%% file: main.tex
\documentclass[conference, 10pt, twoside]{IEEEtran}
\usepackage{spconf}
\usepackage[stretch=80,shrink=80,kerning=true]{microtype}
\input{snippets/tex/paper_abbrev}

\input{snippets/tex/math_abbrev}

\input{snippets/tex/plot_abbrev}

\title{ADMM for ND Line Spectral Estimation using Grid-Free Compressive Sensing from Multiple Measurements with Applications to DOA Estimation}

\makeatletter
\def\@name{S. Semper, F. Römer}
\makeatother

\newcommand\blfootnote[1]{%
  \begingroup
  \renewcommand\thefootnote{}\footnote{#1}%
  \addtocounter{footnote}{-1}%
  \endgroup
}

\address{EMS Research Group, Ilmenau University of Technology}

\pgfplotscreateplotcyclelist{tui_dark_2}{
{thick,tui_orange_dark,mark=*,mark options={thin,scale=1, fill=tui_orange_dark!50!white}},
{thick,tui_green_dark,mark=square*,mark options={thin,scale=1, fill=tui_green_dark!50!white}},
{thick,tui_blue_dark,mark=triangle*,mark options={thin,scale=1, fill=tui_blue_dark!50!white}},
{thick,tui_red_dark,mark=diamond*,mark options={thin,scale=1, fill=tui_red_dark!50!white}},
{thick,black!80!white,mark=otimes*,mark options={thin,scale=1, fill=black!30!white}},
}

\pgfplotsset{
    cycle list name=tui_dark_2,
}

\begin{document}
\ninept
\abovedisplayskip      = 1pt plus 1pt minus 1pt
\abovedisplayshortskip = 1pt plus 1pt minus 1pt
\belowdisplayskip      = 1pt plus 1pt minus 1pt
\belowdisplayshortskip = 1pt plus 1pt minus 1pt

\maketitle
\blfootnote{\emph{S. Semper is funded by DFG under the project ``CoSMoS''. F. Römer is funded by the Fraunhofer Internal Programs under Grant No. Attract 025-601128.}}
\input{tex/abstract.tex}

\section{Introduction and Signal Model}

\input{tex/introduction.tex}

\section{Grid-Free Reconstruction}

\input{tex/reconstruction.tex}

\section{Derivation of the ADMM}

\input{tex/admm.tex}

\section{Numerical Simulations}

\input{tex/numerics.tex}
\bibliographystyle{IEEEtran}
\bibliography{references,snippets/bib/offgrid,snippets/bib/opt,snippets/bib/sigproc,snippets/bib/myown}

\end{document}

%% file: snippets/tex/paper_abbrev.tex
\usepackage{ifthen}
\usepackage[english]{babel}
\usepackage[fleqn]{amsmath}
\usepackage{amsfonts, amssymb, yhmath, bm}
\allowdisplaybreaks
\usepackage{graphicx}
\usepackage{color}
\usepackage[utf8]{inputenc}

\usepackage{psfrag}

\usepackage{enumitem}


%% file: snippets/tex/math_abbrev.tex
\newcommand{\trans}{\mathrm{T}} 
\newcommand{\herm}{\mathrm{H}}

\newcommand{\R}{\bm{\mathrm{R}}}
\newcommand{\C}{\bm{\mathrm{C}}}

\newcommand{\N}{\bm{\mathrm{N}}}

\newcommand{\Real}[1]{{\rm Re}\left\{#1\right\}}
\newcommand{\Imag}[1]{{\rm Im}\left\{#1\right\}}
\newcommand{\Conj}[1]{\overline{#1}}

\newcommand{\ScPr}[2]{{\left\langle #1,#2 \right\rangle}}
\newcommand{\Norm}[1]{{\left\Vert #1\right\Vert}}
\newcommand{\Abs}[1]{{\left| #1 \right|}}
\newcommand{\Text}[1]{{\hspace{3mm} \text{#1} \hspace{3mm}}}

\usepackage{calc}

\newcommand{\Sum}[3]{{\sum\limits_{#1}^{#2}{#3}}}

\newcommand{\D}[0]{\hspace{0.5mm}:\hspace{1.0mm}}

\DeclareMathOperator*{\Argmin}{argmin}

\DeclareMathOperator*{\Min}{min}

\DeclareMathOperator*{\Tr}{tr}

\DeclareMathOperator*{\Vectorize}{vec}

\usepackage{amsthm}
\usepackage{cleveref}

\crefname{Lem}{lemma}{lemmata}

\newtheorem{Thm}{Theorem}[section]
\crefname{Thm}{theorem}{theorems}

\crefname{Pro}{proposition}{propositions}

\crefname{Cor}{corollary}{corollaries}

\theoremstyle{definition}

\crefname{Exm}{example}{examples}

\crefname{Rem}{remark}{remarks}

\crefname{Alg}{algorithm}{algorithms}

\newtheorem{Def}{Definition}[section]
\crefname{Def}{defintion}{defintions}


%% file: snippets/tex/plot_abbrev.tex
\usepackage{tikz}
\usepackage{graphicx}
\usetikzlibrary{matrix,arrows,shapes,trees,fit}
\usetikzlibrary{calc,positioning,scopes,backgrounds}
\newif\ifoutertikznest
\outertikznestfalse
\usetikzlibrary{decorations.markings}

\tikzstyle{lw} = [line width=3pt]

\usepackage{pgfplots}
\pgfplotsset{compat=newest}

\definecolor{box_white}{cmyk}{0.0361,0.0251,0.0166,0}
\definecolor{text_black}{cmyk}{0.7979,0.7417,0.6916,0.6554}
\definecolor{tui_orange_dark}{cmyk}{0,0.6,1,0}
\definecolor{tui_orange_light}{cmyk}{0.0000,0.0876, 0.1474, 0.0157}
\definecolor{tui_green_dark}{cmyk}{1,0,0.5,0.2}
\definecolor{tui_green_light}{cmyk}{0.0576,0.0041, 0.0000, 0.0471}
\definecolor{tui_blue_dark}{cmyk}{1.0000,0.5000,0.0000,0.6000}
\definecolor{tui_blue_light}{cmyk}{0.0920,0.0440,0.0000,0.0196}
\definecolor{tui_red_dark}{cmyk}{0.0000,1.0000,1.0000,0.2000}
\definecolor{tui_red_light}{cmyk}{0.0000,0.1107,0.1107,0.0078}

\pgfplotscreateplotcyclelist{tui_dark}{
{thick,tui_orange_dark,mark=*,mark options={thin,scale=0.75, fill=tui_orange_dark!50!white}},
{thick,tui_green_dark,mark=square*,mark options={thin,scale=0.75, fill=tui_green_dark!50!white}},
{thick,tui_blue_dark,mark=triangle*,mark options={thin,scale=0.75, fill=tui_blue_dark!50!white}},
{thick,tui_red_dark,mark=diamond*,mark options={thin,scale=0.75, fill=tui_red_dark!50!white}},
}

\pgfplotsset{
    cycle list name=tui_dark,
}

\usepackage{datatool}
\usepackage{numprint}
\DTLsetseparator{,}

\definecolor{tui_orange_dark}{cmyk}{0,0.6,1,0}
\definecolor{tui_orange_light}{cmyk}{0.0000,0.0876, 0.1474, 0.0157}
\definecolor{tui_green_dark}{cmyk}{1,0,0.5,0.2}
\definecolor{tui_green_light}{cmyk}{0.0576,0.0041, 0.0000, 0.0471}
\definecolor{tui_blue_dark}{cmyk}{1.0000,0.5000,0.0000,0.6000}
\definecolor{tui_blue_light}{cmyk}{0.0920,0.0440,0.0000,0.0196}
\definecolor{tui_red_dark}{cmyk}{0.0000,1.0000,1.0000,0.2000}
\definecolor{tui_red_light}{cmyk}{0.0000,0.1107,0.1107,0.0078}

%% file: tex/abstract.tex

\begin{abstract}\textbf{
This paper is concerned with estimating unknown multi-dimensional frequencies from linear compressive measurements. This is accomplished by employing the recently proposed atomic norm minimization framework to recover these frequencies under a sparsity prior without imposing any grid restriction on these frequencies. To this end, we give a rigorous derivation of an iterative scheme called alternating direction of multipliers method, which is able to incorporate multiple compressive snapshots from a multi-dimensional superposition of complex harmonics. The key result here is how to formulate the objective function minimized by this scheme and its partial derivatives, which become hard to manage if the dimensionality of the frequencies is larger than $1$. Moreover we demonstrate the performance of this approach in case of $3$D line spectral estimation and $2$D DOA estimation with a synthetic antenna array.}

\emph{Keywords:} Line Spectral Estimation, DOA Estimation, Sparse Recovery, Compressed Sensing, Optimization
\end{abstract}

%% file: tex/introduction.tex

Direction of Arrival (DOA) estimation has been a field of active research for several decades~\cite{KV:96} with a wide range of applications such as radar, sonar, communications, or channel sounding. Conventional techniques either exploit some algebraic structures of the underlying array manifolds or employ iterative solutions of the underlying non-convex maximum likelihood estimation problem \cite{KV:96}.

Later, connections between the DOA estimation problem and the field of compressed sensing (CS) have been discovered \cite{candes2008}. Since the observed signals are sparse in the angular domain, algorithms from the field of sparse signal recovery (SSR) can be applied for DOA estimation \cite{MCW:05,HM:10}. Since the angle is a continuous parameter and its discretization introduces an unwanted model mismatch \cite{ZLG:11}, grid-free SSR methods are of particular interest.

However, up until recently it has not been clear how to derive recovery guarantees for compressive measurements that are more general than just randomly subselecting elements from the samples acquired in the scenario of grid-free multidimensional line spectral estimation. Moreover, these methods only allow to estimate directions of arrival from idealized arrays using measurements from uniform linear antenna arrays.

We have recently shown~\cite{SRHD:18} that the DOA estimation problem with realistic antenna arrays and spatial compression can be reformulated as a line spectral estimation problem~\cite{heckel2017gen_line_spec_est}, where the dimensionality of the frequencies to be estimated is equal to the dimension of the angular domain. Moreover, one can formulate even higher dimensional frequency domain models, where one also aims to estimate Doppler shifts or time delays of arrival. So high dimensional frequency estimation problems with a sparsity prior arise naturally in many fields of signal processing.

A popular approach within this sparsity based framework is to formulate an optimization problem in order to retrieve the unknown frequencies, which is called atomic norm minimization~(ANM). This is a direct generalization of the well studied $\ell^1$ minimization, where very efficient algorithms for exact or approximate solutions exist. ANM itself is an optimization in an infinite dimensional space and as such it is infeasible to solve it directly. Instead one formulates an appropriate dual problem, which is a problem with semidefinite side constraints, for which in theory algorithms with polynomial runtime exist. In the single dimensional case, these algorithms perform well enough, but with increasing dimensions the dual problem becomes infeasibly large for these solvers, so an iterative approach, which delivers reasonable accuracy after a few steps is of high interest.

To this end an, Alternating Direction of Multipliers Method (ADMM) type~\cite{boyd2011admm} of optimization algorithm has already been proposed~\cite{li2016offgridlinespectrumdenoising,bhaskar2013ANDNlinespecest} to solve the one dimensional line spectral estimation problem, but the derivations lack detail and explanation. Moreover, it is unclear how to generalize it to higher dimensions while also incorporating a compression step together with the multiple measurement scenario. Here we address this issue by presenting a general ADMM scheme for multi-dimensional line spectral estimation from multiple compressive measurements and we demonstrate its applicability to 2D DOA estimation.
\subsection{Notation}
The inner product of two matrices $\bm A, \bm B \in \C^{n \times m}$ is defined via $\ScPr{\bm A}{\bm B} = \Tr{\bm A^\herm \bm B}$, the Kronecker product of two matrices $\bm A \in \C^{n \times m}$ and $\bm B \in \C^{k \times l}$ is denoted by $\bm A \otimes \bm B$, $\bm A \diamond \bm B$ on the other hand denotes the column-wise Kronecker product, $\Norm{\bm A}_F$ denotes the Frobenius norm of a matrix $\bm A \in \C^{n \times m}$ and $\Norm{\bm x}_p$ denotes the $\ell^p$ norm for $p \geqslant 1$ and $\bm x \in \C^n$. Moreover, we use the abbreviation for so called multi-indices $[n_1, \dots, n_d] = \{1, \dots, n_1\} \times \dots \times \{1, \dots, n_d\} \subset \N^d$ and $\Re(z)$ and $\Im(z)$ denote the real and imaginary part of a complex number $z \in \C$, whereas $\overline{z} \in \C$ denotes the complex conjugate, which we also use for vectors or higher order tensors element wise. For a given $\bm x \in \C^{n_1 \times \dots \times n_d}$ we define $\Vectorize(\bm x) \in \C^{n_1 \dots n_d}$ as the rearrangement of all elements of $\bm x$ into a single vector.
\subsection{Data Model}
We observe a superposition of complex $d$-dimensional narrow-band harmonics with unknown amplitudes, where $d \geqslant 1$. This results in a model which read as
\[
z_{\bm k}(t)
    = \sum_{i = 1}^r
        s_i(t) \frac{1}{\sqrt{N_1 \cdot \dots \cdot N_d}} \exp\left(-\jmath 2 \pi \ScPr{\bm k}{\bm f_i} \right)
\]
for $\bm k \in [N_1, \dots, N_d]$ with $N_i \geqslant 2$ for $i = 1, \dots, d$ such that $\bm z(t) \in \C^{N_1 \times \dots \times N_d}$. If we define
\begin{align}\label{def_atoms}
\bm a(\bm f)
    &= \bigotimes_{p=1}^d \frac{1}{\sqrt{N_p}} \exp(-\jmath 2 \pi k_p f_p)\notag\\
    &= \frac{1}{\sqrt{N_1 \dots N_d}} \Vectorize\left(\left[\exp\left(-\jmath 2 \pi \ScPr{\bm k}{\bm f} \right)\right]_{\bm k \in [N_1, \dots, N_d]}\right)
\end{align}
so that $\bm a(\bm f) \in \C^{M \times 1}$ we get
\begin{align}\label{signal_model}
    \Vectorize \bm z(t) = \sum_{i = 1}^r s_i(t) \bm a(\bm f_i).
\end{align}
If we now collect multiple measurements at $K$ points in time $z(t_k)$, we can write
\[
\bm Z = [\Vectorize \bm z(t_1), \dots,\Vectorize \bm z(t_K)] = \bm A(\bm f_1, \dots, \bm f_r) \bm S(t_1, \dots t_k)
\]
for $\bm Z \in \C^{M \times K}$, $\bm A(\bm f_1, \dots, \bm f_r) \in \C^{M \times r}$ ans $\bm S \in \C^{r \times K}$, where $M = N_1 \cdot \dots \cdot N_d$. In order to model a compressive sensing scenario, we assume that in stead of observing $\bm Z$ directly, we model our observations as a collection of linear measurements applied to $\bm Z$. These are represented by a combining matrix $\bm \Phi \in \C^{m \times M}$ for some $m \leqslant M$ and thus our final model reads as
\begin{align}\label{linear_compr_model}
    \bm Y = \bm \Phi \bm Z + \bm N = \bm \Phi \bm A \bm S + \bm N,
\end{align}
where $\bm N \in \C^{m \times K}$ accounts for additive measurement noise. For this scenario it has already been shown in~\cite{heckel2017gen_line_spec_est} that sparse recovery techniques like ANM can recover the underlying frequencies from compressive measurements with high probability when imposing a separation condition and in case of a suitably generated compression matrix $\bm \Phi$, i.e. where the entries are drawn i.i.d. from a sub-Gaussian distribution.
\subsection{Application to DOA estimation}
The above model for frequency estimation can be used to cast the problem of $2$-d DOA estimation with realistic antenna elements. To this end, let $\bm{r}(\bm \theta) \D [0, 2 \pi ) \times [0, \pi] \rightarrow \C^M$ model the response of an array comprising of $M$ antennas for a planar wave impinging from azimuth and elevation angle $\bm \theta = (\theta_1, \theta_2)$. Naturally, each element $r_m(\bm \theta)$, $m=1, \ldots, M$ is a periodic function in $\theta_1$ and most importantly can be measured in practice for a specific antenna array geometry. Moreover, since beam patterns are typically quite smooth functions, they can be very well approximated by a truncated Fourier series \cite{LRT:04} given by
\begin{align}
    r_m(\bm \theta) \approx \frac{1}{\sqrt{L_1 L_2}}
    \sum_{\ell_1 = -\frac{L_1-1}{2}}^{\frac{L_1-1}{2}}
    \sum_{\ell_2 = -\frac{L_2-1}{2}}^{\frac{L_2-1}{2}}
    g_{m, \ell_1, \ell_2} {\rm e}^{\jmath (\theta_1 \ell_1 + \theta_2 \ell_2)}, \label{eqn:eadf_scalar}
\end{align}
where we have considered odd numbers of $L_1$ and $L_2$ terms respectively. Rewriting above formula in matrix form yields
\[r_m(\bm\theta) \approx \bm a(\theta_1)^\trans \cdot \bm G_m \cdot \bm a(\theta_2) = \Vectorize(\bm G_m) \cdot (\bm a(\theta_1) \otimes \bm a(\theta_2)),\]
where the matrix $\bm G_m$ collects the Fourier coefficients of $r_m(\bm \theta)$. If we now observe $S$ plain narrowband waves from the far field of the array with unknown time varying amplitudes $s_1(t), \dots s_S(t)$ and unknown directions of arrival $\bm \theta_1, \dots, \bm\theta_S$ and define
\[
\bm G = \begin{bmatrix}
    \Vectorize(\bm G_1)^\trans, \dots, \Vectorize(\bm G_m)^\trans
\end{bmatrix}^\trans \Text{and} \bm A = \bm A(\bm\theta_{1}, \dots, \bm\theta_{S})
\]
we arrive at the following model for the measurements we collect at the $M$ ports of the antenna array
\[
\bm z(t) = \bm G \bm A \bm s(t) \in \C^{M \times 1}.
\]
If we now also take snapshots $\bm z(t_1), \dots, \bm z(t_K)$ for $K \in \N$, apply spatial compression as was recently proposed in~\cite{SRHD:18} for the one-dimensional case, we get
\begin{align}
    \bm Y = \bm \Psi \bm Z + \bm N = \bm \Psi \bm G \bm A \bm S + \bm N,
    \label{doa_lse_model}
\end{align}
where again $\bm N$ accounts for additive measurement noise. So we arrive at a $2$-dimensional line spectral estimation problem from multiple compressed snapshots as in \eqref{linear_compr_model} if we set $\bm \Phi = \bm \Psi \bm G$ and any technique to recover the unknown frequencies in \eqref{linear_compr_model} is suitable to recover the unknown directions of arrival. This task and means to tackle it by using a sparse recovery approach are subject of the next section.

%% file: tex/reconstruction.tex

In the recent literature~\cite{angel2015comp_beamform,TBSR:13} it has been put forward to use ANM as a generalization of the conventional $\ell^1$-norm minimization in compressive sensing when one aims at recovering parameters, which are not constrained to be resided on a discrete grid but rather on a continuous manifold. The next section introduces the notion of the atomic norm in the multiple snapshot $d$-dimensional line spectral estimation scenario and the corresponding optimization problem for estimating these unknown frequencies from compressive measurements.
\begin{Def}[atomic norm]
Let $\mathcal{A} \subset \C^N$ be an arbitrary set. Then, the \textbf{atomic norm} $\Norm{\cdot}_\mathcal{A} \D \C^N \rightarrow \R^+_0$ is defined as
\begin{align}\label{atomic_norm}
\bm X \mapsto \Norm{\bm X}_\mathcal{A} = \inf\left\{
    \sum_{i = 1}^r \Abs{c_i} \middle\vert
    \bm X = \Sum{i = 1}{r}{c_i \bm a_i,\quad \bm a_i \in \mathcal{A}}
\right\}.
\end{align}
\end{Def}
As it turns out, this is a direct generalization of the conventional $\ell^1$ norm $\Norm{\bm u}_\mathcal{A} = \Norm{\bm x}_1$ of a vector $\bm u = \bm A \bm x$ if one takes $\mathcal{A}$ to be the set of columns of $\bm A$. Next, following~\cite{lichi2016admm_mmv} we specialize the general atomic set from above such that it fits the structure of $d$-dimensional line spectral estimation from $K$ measurements:
\begin{Def}[atomic set]\label{lse_atoms}
Let $\bm a \D (0,1]^d \rightarrow \C^{N_1 \times \dots \times N_d}$ be defined as in \eqref{def_atoms} then the \textbf{atomic set} for $K$ snapshots is defined as
\begin{align*}
\mathcal{A} = \left\{
    \bm a(\bm f) \bm b^\herm \middle\vert
    \bm b \in \C^K, \Norm{\bm b}_2 = 1,
    \bm f \in (0, 1]^d
\right\} \subset \C^{N_1 \dots N_d \times K}.
\end{align*}
\end{Def}
In some way the above set captures the model in the sense that it represent all possibly occuring harmonics in the multiple snapshot case, so our signal in \eqref{signal_model} is comprised of a sparse and linear superposition of elements in above atomic set. To extract the parameters corresponding to these atoms we aim at solving a grid free sparse recovery problem, where we minimize the atomic norm for the above defined set $\mathcal{A}$.

So, the problem for atomic norm minimization as proposed in~\cite{bhaskar2013ANDNlinespecest} and extended to compressive measurements in \cite{SRHD:18} reads as
\[
\Min_{\bm Z \in \C^{M \times K}} \Norm{\bm Z}_\mathcal{A} \Text{subject to} \Norm{\bm Y - \bm \Phi \bm Z}_F^2 \leqslant \varepsilon,
\]
for suitably chosen $\varepsilon$, where we additionally employ a compression matrix $\bm \Phi \in \C^{m \times M}$ as before, which we can choose freely. Now, still following \cite{bhaskar2013ANDNlinespecest} and \cite{lichi2016admm_mmv} we aim at posing the equivalent dual problem, since directly minimizing the atomic norm is an infinite dimensional optimization problem. For this we need the concept of specially structured matrices, which are defined as follows.
\begin{Def}[multilevel Toeplitz matrices]
Let for $d \in \N$ $\bm u \in \C^{n_1 \times 2 \cdot n_2 - 1 \times \dots \times 2 \cdot n_d - 1}$ be the tensor of defining elements and $\bm n \in [n_1, \dots, n_d]$ be a vector of dimension sizes. Now, the \textbf{Hermitian $\bm d$-level Toeplitz matrix} $\bm{T^\herm}_{(\bm{n},d)}(\bm u)$ is recursively defined with a blockwise Toeplitz structure as
\[
\bm{T^\herm}_{(\bm{n},d)}(\bm u) =
\begingroup
\setlength\arraycolsep{0pt}
\begin{bmatrix}
    \bm{T^\herm}_{(\bm m, \ell)}(\bm{u}_{[1, \bm m]}) &
    \dots &
    \bm{T}_{(\bm m, \ell)}(\bm{u}_{[n_1, \bm m]}) \\

    \vdots &
    \ddots &
    \vdots \\

    \bm{T}_{(\bm m, \ell)}(\bm{u}_{[n_1, \bm m]})^\herm &
    \dots &
    \bm{T^\herm}_{(\bm m, \ell)}(\bm{u}_{[1, \bm m]}) \\
\end{bmatrix}
\endgroup
\]
where $\bm m = [n_2,\dots, n_d]$ and $\ell = d-1$. Moreover a \textbf{$\bm d$-level Toeplitz matrix} is defined with above notation and again $\bm u \in \C^{2 \cdot n_1 - 1 \times 2 \cdot n_2 - 1 \times \dots \times 2 \cdot n_d - 1}$ being the tensor of defining elements via
\[
\bm{T}_{(\bm{n},d)}(\bm u) =
\begingroup
\setlength\arraycolsep{0pt}
\begin{bmatrix}
    \bm{T}_{(\bm m, \ell)}(\bm{u}_{[1, \bm m]}) &
    \dots &
    \bm{T}_{(\bm m, \ell)}(\bm{u}_{[2 \cdot n_1 - 1, \bm m]}) \\

    \vdots &
    \ddots &
    \vdots \\

    \bm{T}_{(\bm m, \ell)}(\bm{u}_{[n_1, \bm m]}) &
    \dots &
    \bm{T}_{(\bm m, \ell)}(\bm{u}_{[1, \bm m]}). \\
\end{bmatrix}
\endgroup
\]
\end{Def}
With these structured matrices at hand, we can reformulate the calculation of the atomic norm in the line spectral estimation case.
\begin{Thm}[\cite{candes2013superresnoisydata}, \cite{TBSR:13}]
With $\mathcal{A}$ given as in Definition \ref{lse_atoms} the following equality holds:
\begin{align}
\Norm{\bm X}_\mathcal{A} = & \min_{\bm W,\bm u}
    \Tr{\bm{T^\herm}_{(\bm{n},d)}(\bm u)} + \Tr \bm W \\
    & \text{subject to} \quad
    \begin{bmatrix}
        \bm{T^\herm}_{(\bm{n},d)}(\bm u) & \bm X \\
        \bm X^\herm & \bm W
    \end{bmatrix} \succeq \bm 0  \notag.
\end{align}
\end{Thm}
The above theorem transforms the infinite dimensional problem of  calculating the atomic norm of multiple snapshots $\bm X$ into a semidefinite optimization program, which can be solved efficiently in theory. With this result at hand, we can now pose the dual problem of atomic norm minimization for generalized line spectral estimation with the model in \eqref{linear_compr_model} via
\begin{align}
    &\min_{\bm W, \bm u, \bm X} \Tr{\bm{T^\herm}_{(\bm{n},d)}(\bm u)} + \Tr{\bm W} \label{dual_prob_anm} \\
		&\text{subject to} \quad
        \begin{bmatrix}
            \bm{T^\herm}_{(\bm{n},d)}(\bm u) & \bm Z \\
            \bm Z^\herm & \bm W
        \end{bmatrix} \succeq \bm 0, \Norm{\bm Y - \bm \Phi \bm Z}^2_F \leqslant \varepsilon  \notag,
\end{align}
which can for instance be found in \cite{yang2016MLtoeplitz}. The key point here is now that the resulting $\bm{T^\herm}_{(\bm{n},d)}(\bm u^*)$ for a dual optimal $\bm u^*$ is an estimate of the covariance of the underlying signal. So the final step would be to apply any covariance based spectral estimator, like MUSIC, ESPRIT or a Vandermonde decomposition to $\bm{T^\herm}_{(\bm{n},d)}(\bm u^*)$.

Although we succeeded in reducing the original problem complexity significantly, for larger dimensions, so $d \geqslant 2$ the explicit solvers for semidefinite programs take prohibitively many iterations to deliver feasible results if one is able to provide it with the side constraints in a reasonable way at all. So a more direct and iterative approach would be advantageous, which is also specifically tailored to the semidefinite program at hand. To this end, we aim at providing an ADMM type algorithm, which approximates a solution to above problem reasonably well and reasonably fast.

%% file: tex/admm.tex

Next, we formulate the iterative update steps of the ADMM in order to approximate a solution to \eqref{dual_prob_anm}. To this end and following \cite{boyd2011admm}, we make use of the augmented Lagrangian of the problem in \eqref{linear_compr_model}, which reads as
\begin{align}
   \min_{\bm W,\bm u, \bm Z, \bm V \succeq \bm 0, \bm \Lambda \succeq \bm 0} & \mathcal L(\bm W,\bm u, \bm Z, \bm V , \bm \Lambda ) \label{ML_LAG_ADMM}
\end{align}
where $\mathcal{L}(\bm W, \bm u, \bm Z, \bm V, \bm \Lambda) \D \C^{K \times K} \times \C^{N_1 \times 2 N_2 - 1 \dots \times 2 N_d - 1} \times \C^{M \times K} \times \C^{M + K \times M + K} \times \C^{M + K \times M + K} \rightarrow \R$ and its values are defined by
\begin{align*}
\mathcal L(\bm W, &\bm u, \bm Z, \bm V, \bm \Lambda ) = \left\langle
    \bm \Lambda, \bm V
    - \underbrace{
        \begin{bmatrix}
            \bm{T^\herm}_{(\bm{n},d)}(\bm u) & \bm Z \\
            \bm Z^\herm & \bm W
        \end{bmatrix}}_{= \bm T}
    \right\rangle \\
    &+ \frac 12 \Norm{\bm \Phi \bm Z - \bm Y}^2_2 + \frac{\tau}{2}(\Tr\bm W + \Tr\bm{T^\herm}_{(\bm{n},d)}(\bm u)) \\
    &+ \frac{\rho}{2} \Norm{
    \bm V
    - \begin{bmatrix}
        \bm{T^\herm}_{(\bm{n}, d)}(\bm u) & \bm Z \\
        \bm Z^\herm & \bm W
    \end{bmatrix}}^2_F,
\end{align*}
where $\tau > 0$ and $\rho > 0$ are suitably chosen constants. Here $\tau$ plays the role of a regularizing parameter between data fitting and the magnitude of the atomic norm of $\bm Z$. We also partition the matrices $\bm \Lambda$ and $\bm V$ such that they match the partitioning of the blocks in $\bm T$:
\[
\bm \Lambda = \begin{bmatrix}
\bm{\hat{\Lambda}} & \bm \Lambda_1 \\
\bm \Lambda_1 & \bm \Lambda_0
\end{bmatrix}
\quad\mbox{and}\quad
\bm V = \begin{bmatrix}
\bm{\hat{V}} & \bm V_1 \\
\bm V_1 & \bm V_0
\end{bmatrix}.
\]
The objective of the next paragraphs is to calculate the partial derivatives of $\mathcal{L}$, which has to be handled carefully, since we have to consider the fact that the variables $\mathcal{L}$ depends on are complex valued and highly structured.
\subsection{Wirtinger Calculus}
To calculate the partial derivatives of functions which depend on complex variables and map to $\R$ we make use of the so called Wirtinger calculus. To this end, let $f$ be a function $f: \C^n \mapsto \R$, then its Wirtinger derivative is defined as
\[
\frac{\partial f}{\partial \bm x} = \frac 12 \left( \frac{\partial f}{\partial \bm y} - \jmath \frac{\partial f}{\partial \bm z} \right)
\quad\mbox{and}\quad
\frac{\partial f}{\partial \Conj{\bm x}} = \frac 12 \left( \frac{\partial f}{\partial \bm y} + \jmath \frac{\partial f}{\partial \bm z} \right),
\]
where $\bm y = \Real{\bm x}$ and $\bm z = \Imag{\bm x}$.
Since we have specifically structured functions and matrices in $\mathcal{L}$, we need only the following three simple rules
\begin{align}
&\frac{\partial \langle \bm b, \bm a\rangle}{\partial \Conj{\bm b}} =
    \frac{\partial \bm{b}^\herm \bm a}{\partial \Conj{\bm b}} = \bm a \mbox{,}\quad
\frac{\partial \langle \bm a, \bm b\rangle}{\partial \Conj{\bm b}} =
    \frac{\partial \bm{a}^\herm \bm b}{\partial \Conj{\bm b}} = \bm 0 \label{eqn:wder_lin_c} \\
\mbox{and}\quad
&\frac{\partial \Norm{\bm{A} \bm{b} - \bm{c}}^2}{\partial \Conj{\bm{b}}} =
    \bm A^\herm (\bm A \bm b - \bm c). \label{eqn:wder_norm}
\end{align}
All three of them can easily be extended to the case where $\bm a$ and $\bm b$ are matrices, since both the inner product of matrices we use here and the Frobenius norm induced by it treat matrices as if they were vectors realigned into matrices.
\subsection{Special Derivatives}
Taking a close look at $\mathcal{L}$ we see that the only derivatives which are not straightforward to calculate are those with respect to $\bm u$, since it is the defining tensor of the multilevel Toeplitz structure. In this case the expressions of interest are
\[
\frac{\partial}{\partial \bm u} \langle \bm A, \bm{T^\herm}_{(\bm{n},d)}(\bm u) \rangle
\quad\mbox{and}\quad
\frac{\partial}{\partial u_{\bm e}}
\langle
    \bm{T^\herm}_{(\bm{n},d)}(\bm u),
    \bm{T^\herm}_{(\bm{n},d)}(\bm u)
\rangle
\]
for a given Hermitian matrix $\bm A$. To this end, for given $n \in \N$ and $p \in [n] \cup -[n] \cup \{0\}$, we define $\bm{S}_{n}^p$ via
\[
    \left[\bm{S}_n^p\right]_{(k,\ell)} = 1 \Text{for} \ell-k= p - n.
\]
Note that $\bm{S}_n^p$ has $1$ only on a shifted diagonal and $\bm S_n^0 = \bm I_n$. Now, we can rewrite
\begin{align}
\bm{T^\herm}_{(\bm{n},d)}(\bm u) = \bm T_{\mathrm{upper}} + \bm T_{\mathrm{lower}}
\end{align}
where $\bm T_{\mathrm{upper}}$ and $\bm T_{\mathrm{lower}}$ are constructed by explicitly unraveling the recursive definition of $\bm{T^\herm}_{(\bm{n},d)}(\bm u)$ while keeping the Hermitian symmetry in mind. So they are defined as
\begin{align}
\bm T_{\mathrm{upper}} &= \sum\limits_{\bm i \in \mathcal{N}} \left(\bm S_{N_1}^{i_1 - 1} \otimes \bm S_{N_2}^{i_2 - N_2} \otimes \dots \otimes \bm S_{N_d}^{i_d - N_d}\right) u_{\bm i} \\
\bm T_{\mathrm{lower}} &= \bm T_{\mathrm{upper}}^\herm,
\end{align}
where $\mathcal{N} = [N_1, 2 N_2 - 1,  \dots, 2 N_d - 1]$. Next, we calculate for given multi-index $\bm i \in \mathcal{N}$
\begin{align}
&\frac{\partial}{\partial u_{\bm i}} \langle \bm A, \bm{T^\herm}_{(\bm{n},d)}(\bm u) \rangle =\\
    = &\left\langle
       \bm A,
       \bm S_{N_1}^{i_1 - 1} \otimes \bm S_{N_2}^{i_2 - N_2} \otimes \dots \otimes \bm S_{N_d}^{i_d - N_d}
       \right\rangle.
\end{align}
For a shorter notation we define the operator $\mathfrak{D}_{\bm n, d} : \C^{M \times M} \rightarrow \C^{N_1 \times 2 N_2 - 1\dots \times 2 N_d - 1}$ via
\begin{align*}
    \bm A &\mapsto \mathfrak{D}_{\bm n, d}(\bm A)
    = \left(\frac{\partial}{\partial u_{\bm i}}\langle \bm A, \bm T_{\bm n, d}(\bm u)\rangle\right)_{\bm i \in \mathcal{N}} = \\
    &=
    \left(\left\langle
    \bm A,
    \bm S_{N_1}^{i_1 - 1} \otimes \bm S_{N_2}^{i_2 - N_2} \otimes \dots \otimes \bm S_{N_d}^{i_d - N_d}
    \right\rangle\right)_{\bm i \in \mathcal{N}}.
\end{align*}
This operator results in a tensor with the same dimensions as $\bm u$ and each entry at position $\bm i \in \mathcal{N}$ represents the sum of the elements in $\bm A$ which occur at the same position as $\bm u_{\bm i}$ in $\bm{T^\herm}_{(\bm{n},d)}(\bm u)$.

Now for some $\bm i \in \mathcal{N}$ and $\bm u = \bm v + \jmath \bm w$ we can also calculate (Note that here we identify the multi-index $\bm i$ with the the tensor of order $d$ which has zeros everywhere except a single $1$ at position $\bm i$.):
\begin{align*}
    &\frac{\partial}{\partial u_{\bm i}}
    \langle
        \bm{T^\herm}_{(\bm{n},d)}(\bm u),
        \bm{T^\herm}_{(\bm{n},d)}(\bm u)
    \rangle
    =
    \frac{\partial}{\partial u_{\bm i}}
    \Norm{\bm{T^\herm}_{(\bm{n},d)}(\bm u)}_F^2
    = \\
    & = \frac{\partial}{\partial u_{\bm i}}
    \left(
    \sum\limits_{\bm i^\prime \in \mathcal{N}} \left[
    \Norm{\bm{T^\herm}_{(\bm{n},d)}(\bm i^{\prime})}_F^2
            v_{\bm i^{\prime}}^2
    - \Norm{\bm{T^\herm}_{(\bm{n},d)}(\bm i^{\prime})}_F^2
            w_{\bm i^{\prime}}^2
    \right]
    \right) = \\
    & = 2 f_{\bm n}(\bm i) \overline{u}_{\bm i},
\end{align*}
where $f_{\bm n}(\bm i)$ represents the number of occurrences of $u_{\bm i}$ in the Hermitian multilevel Toeplitz matrix $\bm{T^\herm}_{(\bm{n},d)}(\bm u)$.

With this intuition at hand, we can easily infer that $\left(f_{\bm n}(\bm i)\right)_{\bm i \in \mathcal{N}} = \mathfrak{D}_{\bm n, d}(\bm{\mathrm{1}})$, where $\bm{\mathrm{1}} \in \C^{M \times m}$ is a matrix with all entries equal to $1$. Now, these rules together with \eqref{eqn:wder_lin_c} and \eqref{eqn:wder_norm} can be used to establish the following results about the partial derivatives of $\mathcal{L}$ which read as:
\begin{align}
\frac{\partial \mathcal L}{\partial \bm W} &= \frac{\tau}{2}\bm I_K - \bm \Lambda_0 - \rho (\bm V_0 - \bm W) \label{deriv_W}, \\
\frac{\partial \mathcal L}{\partial \bm u} &= \frac{\tau}{2} \bm i_1 - \mathfrak{D}_{\bm n, d}(\bm{\hat{\Lambda}}) + \frac{\rho}{2} \left(\mathfrak{D}_{\bm n, d}(\bm{1}) - 2\mathfrak{D}_{\bm n, d}(\bm{\hat{V}})\right) \label{deriv_u}, \\
\frac{\partial \mathcal L}{\partial \overline{\bm Z}} &=
\frac 12 (\bm \Phi^\herm \bm \Phi \bm Z - \bm \Phi^\herm \bm Y) - \bm{\hat{\Lambda}} - \rho \left(\bm{\hat{V}} - \bm Z\right)\label{deriv_Z},
\end{align}
where $\bm i_1$ is the tensor of the same dimension as $\bm u$ with entries all equal to $0$ except at the position of $u_{[1, \dots, 1]}$. With these three derivatives at hand we can proceed to formulate the explicit update steps for the ADMM iteration.
\subsection{Update Steps}
This section gives the explicit updates rules for the alternating updates of the ADMM. In general and according to \cite{bhaskar2013ANDNlinespecest} the iteration after step $k$ can be expressed as
\[
(\bm W^{k+1}, \bm u^{k+1}, \bm Z^{k+1}) \leftarrow \Argmin\limits_{\bm W, \bm u, \bm Z} \mathcal{L}(\bm W, \bm u, \bm Z, \bm V^k, \bm \Lambda^k)
\]
\[
\bm V^{k+1} \leftarrow \Argmin\limits_{\bm V \succeq 0} \mathcal{L}(\bm W^{k+1}, \bm u^{k+1}, \bm Z^{k+1}, \bm V, \bm \Lambda^k)
\]
and
\[
\bm \Lambda^{k+1} = \bm \Lambda^{k} + \rho (\bm V^{k+1} - \bm T^{k+1})
\]
By using the derivatives of the previous section in equations \eqref{deriv_W}, \eqref{deriv_u} and \eqref{deriv_Z} we can compute above updates in closed form by setting the respective derivatives to $0$, thus satisfying the necessary conditions for optimality, and solving for the variable of interest. Finally, they read as follows
\begin{align}
\bm u^{k+1} &= \frac{1}{\rho \mathfrak{D}_{\bm n, d}(\bm{\mathrm{1}})}\left(
\mathfrak{D}_{\bm n, d}(\bm{\hat{\Lambda}}^k - \rho \bm{\hat{V}}^k) - \frac{\tau}{2} \bm i_{1}
\right) \label{update_u}\\
\bm W^{k+1} &= \frac{1}{\rho} \bm \Lambda_0^k + \bm V_0^k - \bm I \frac{\tau}{2\rho} \label{update_W}\\
\bm Z^{k+1} &= \left(\bm \Phi^\herm \bm \Phi - \rho \bm I\right)^{-1} \cdot \left(\bm \Phi \bm Y + \bm \Lambda_1^k + \rho \bm V_1^k\right),\label{update_Z}
\end{align}
where $\left(\bm \Phi^\herm \bm \Phi - \rho \bm I\right)^{-1}$ can be precomputed in advance to avoid repeatedly solving a linear system. Furthermore we also update $\bm T$ iteratively from the current variables according to
\begin{align}
\bm T^{k+1} = \begin{bmatrix}
    \bm{T^\herm}_{(\bm{n},d)}(\bm u^{k+1}) & \bm Z^{k+1} \\
    {\bm Z^\herm}^{k+1} & \bm W^{k+1}
\end{bmatrix}\label{update_T}
\end{align}
whereas we finally can also update $\bm V$ via
\begin{align}
\bm V^{k+1} = \bm{\mathfrak{P^\succeq}}\left(\bm T^{k+1} - \rho \bm \Lambda^k\right),\label{update_V}
\end{align}
where $\bm{\mathfrak{P^\succeq}}$ is the orthogonal projection onto the positive cone of the positive semi-definite matrices, which can be realized numerically by an eigendecomposition and is the computationally most expensive step during the iterate updates. With this we have completed the iterations for ADMM which now only consists of initializing $(\bm W^0, \bm Z^0, \bm u^0, \bm V^0, \bm \Lambda^0)$ and iteratively carrying out the steps in \eqref{update_u}, \eqref{update_W}, \eqref{update_Z}, \eqref{update_T} and \eqref{update_V}. With the operators defined as above, which can be implemented recursively for arbitrary dimensions, one can create a very general implementation as well. Next, we proceed with the numerical evaluation.

%% file: tex/numerics.tex

To quantify the performance of our estimator, we compare it to the deterministic Cramér-Rao Bound (CRB). With the proposed spatial compression, the deterministic CRB for the $d$-dimensional case with $K$ snapshots can be computed via~\cite{IRLKRLGDT:17}
\begin{align}
 C(\bm \theta) = \frac{\sigma^2}{2 \cdot K} \Tr\left(\left[ \Re(\bm D^\herm \bm \Pi_{\bm{G}}^\perp \bm D \odot (\bm{\mbox{1}}_{d \times d} \otimes \hat{\bm R})^\trans) \right]^{-1} \right), \label{eqn:crb_uncomp}
\end{align}
with $\bm \Pi_{\bm{G}}^\perp = \bm I - \bm G (\bm G^\herm \bm G)^{-1}\bm G^\herm$ and $\hat{\bm R} = 1/K \cdot \bm S \bm S^\herm$ being the sample covariance and we have set
\begin{align*}
\bm G &= \bm\Phi \bm A(\bm \theta_1, \dots, \bm \theta_S),~ \bm D_i = \frac{\partial}{\partial \bm \theta_i} \bm G,~ \bm D = \left[\bm D_1, \dots, \bm D_d \right]
\end{align*}

First, we carry out the atomic norm minimization for 3D line spectral estimation with uncompressed measurements, so $\bm \Phi = \bm I$ in \eqref{linear_compr_model} and also compressed measurements, where the entries of $\bm \Phi$ are drawn i.i.d. from a zero-mean Gaussian distribution and then we project the columns to the complex unit sphere in $\C^m$, thus normalizing the columns independently. Here, we chose $m = \lfloor \rho \cdot M \rfloor$ according to some compression rate $\rho \in (0, 1]$. In case of $\bm \Phi = \bm I$, so $\rho = 1$, we also run 3D-Standard-ESPRIT~\cite{roy1989esprit} directly on $\bm Y$ as a comparison, which is only applicable in this case, since ESPRIT is not able to deal with compressive measurements of the kind employed here. In any case, we choose $\tau = \sigma^{0.8}$ and $\rho = 0.05$ to run the ADMM and we initialized the state variables $(\bm W^0, \bm u^0, \bm Z^0, \bm V^0, \bm \Lambda^0)$ by sampling the real and imaginary parts from standard Gaussian distributions.

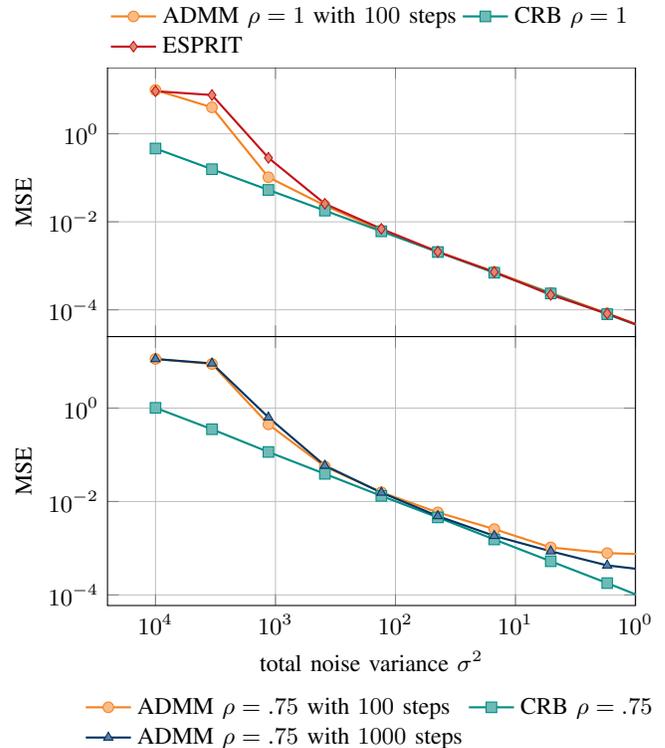
\begin{figure}
\vspace{-3mm}
\begin{tikzpicture}
\begin{semilogyaxis}[
    grid=both,
    grid style={line width=.1pt, draw=gray!10},
    major grid style={line width=.2pt,draw=gray!50},
    name = top,
    legend style = {
        at={(0.5, 1.0)},
        anchor=south,
        legend columns=2,
        draw=none,
        legend cell align={left},
    },
    ylabel = {MSE},
    height = 0.6\linewidth,
    width = \linewidth,
    xmax = 0,
    xticklabel=\empty,
]
\addplot table[col sep = comma, x = snr, y = admm_mean]
    {simulation/lse_C100_S3_st100_K100_D3_3_3_3.csv};
\addplot table[col sep = comma, x = snr, y = crb_mean]
    {simulation/lse_C100_S3_st100_K100_D3_3_3_3.csv};
\addplot[thick,tui_red_dark,mark=diamond*,mark options={thin,scale=1, fill=tui_red_dark!50!white}] table[col sep = comma, x = snr, y = esprit_mean]
    {simulation/lse_C100_S3_st100_K100_D3_3_3_3.csv};
\legend{ADMM $\rho = 1$ with $100$ steps, CRB $\rho = 1$, ESPRIT}
\end{semilogyaxis}
\begin{semilogyaxis}[
    grid=both,
    grid style={line width=.0pt, draw=none},
    major grid style={line width=.2pt,draw=gray!50},
    minor tick num=0,
    at=(top.south),
    anchor=north,
    legend style = {
        at={(0.5, -0.3)},
        anchor=north,
        legend columns=2,
        draw=none,
        fill=none,
        legend cell align={left},
    },
    ylabel = {MSE},
    xlabel = {total noise variance $\sigma^2$},
    height = 0.6\linewidth,
    width = \linewidth,
    xmax = 0,
    xticklabel={$10^{
    \pgfmathparse{-1 * \tick / 10}
    \pgfmathprintnumber[precision=2]{\pgfmathresult}}$}
]
\addplot table[col sep = comma, x = snr, y = admm_mean]
    {simulation/lse_C75_S3_st100_K100_D3_3_3_3.csv};
\addplot table[col sep = comma, x = snr, y = crb_median]
    {simulation/lse_C75_S3_st100_K100_D3_3_3_3.csv};
\addplot table[col sep = comma, x = snr, y = admm_mean]
    {simulation/lse_C75_S3_st1000_K100_D3_3_3_3.csv};
\legend{ADMM $\rho = .75$ with $100$ steps, CRB $\rho = .75$, ADMM $\rho = .75$ with $1000$ steps}
\end{semilogyaxis}
\end{tikzpicture}
\vspace{-8mm}\caption{ADMM reconstruction performance for line spectral estimation of $S = 3$ sources and $d = 3$-dimensional frequencies with $\bm k \in [3,3,3]$ in comparison to standard ESPRIT using $K = 100$ snapshots.}\label{lse_perf}
\vspace{-5mm}
\end{figure}

The results in Figure \ref{lse_perf}, where we plot the reconstruction error versus the noise variance $\sigma^2$, show that the derived ADMM approach is able to replicate the performance predicted by the CRB for the case $\rho = 1$, thus delivering the same performance as 3D-Standard-ESPRIT.

In the case $\rho = 0.75$ we see that the ADMM algorithm's performance highly depends on the number of steps carried out to estimate the covariance $\bm{T^\herm}_{(\bm{n},d)}(\bm u^*)$, since the error floor decreases when iterating for $1000$ steps instead of $100$. In conclusion, this means that the ADMM approach also achieves the CRB after a suitable amount of iterations.

\begin{figure}
\begin{tikzpicture}
\begin{axis}[
    grid=both,
    grid style={line width=.0pt, draw=none},
    major grid style={line width=.2pt,draw=gray!50},
    height = 0.5\linewidth,
    width = \linewidth,
    ymin = -1,
    ymax = +1,
    xlabel = {azimuth [rad]},
    ylabel = {elevation [rad]}
    ]
\addplot+[only marks] table[col sep = comma, x = t1, y = t2]
    {simulation/2DDOA_31.csv};
\addplot+[only marks, opacity=0.5] table[col sep = comma, x = e1, y = e2]
    {simulation/2DDOA_31.csv};
\legend{true locations, estimated locations}
\end{axis}
\end{tikzpicture}
\vspace{-3mm}\caption{Simulation results from a 2D DOA estimation scenario using a stacked uniform linear array with $3$ stacks and $12$ elements each with noise standard deviation of $\sigma = 0.01$.}\label{doa_perf}
\vspace{-5mm}
\end{figure}
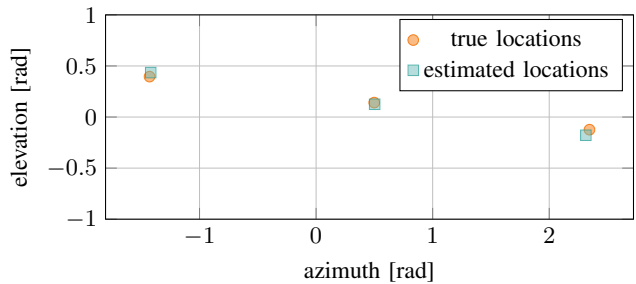

Moreover, we use the derived algorithm's flexibility and apply it to the 2D DOA estimation problem with a $12 \times 3$ stacked circular array, where the stacks are aligned in the $x$-$y$-plane with distance $\mathrm{d}z = 0.375 \lambda$ and diameter $12 / 16 \lambda = 0.75 \lambda$ and $\lambda$ is the wavelength of the impinging wave. We use the Fourier coefficients of this (synthetic) array to formulate the DOA problem into a line spectral estimation problem as in \eqref{doa_lse_model} for $\bm \Psi = \bm I$. A single scenario is depicted in Figure \ref{doa_perf} where the noise variance is $\sigma^2 = 0.001$ and we recover these locations in the 2D angular domain from $K = 100$ snapshots. It is worth noting that these results can also be obtained from realistic arrays described by measured data and also if $\bm \Psi$ actually carries out a compression step.

Summarizing, we have demonstrated that the derived ADMM iteration scheme is capable of recovering the unknown multidimensional frequencies from a very general model and how it can be applied to 2D DOA estimation using arbitrary antenna arrays.

%% file: main.bbl
\begin{thebibliography}{10}
\providecommand{\url}[1]{#1}
\csname url@samestyle\endcsname
\providecommand{\newblock}{\relax}
\providecommand{\bibinfo}[2]{#2}
\providecommand{\BIBentrySTDinterwordspacing}{\spaceskip=0pt\relax}
\providecommand{\BIBentryALTinterwordstretchfactor}{4}
\providecommand{\BIBentryALTinterwordspacing}{\spaceskip=\fontdimen2\font plus
\BIBentryALTinterwordstretchfactor\fontdimen3\font minus
  \fontdimen4\font\relax}
\providecommand{\BIBforeignlanguage}[2]{{%
\expandafter\ifx\csname l@#1\endcsname\relax
\typeout{** WARNING: IEEEtran.bst: No hyphenation pattern has been}%
\typeout{** loaded for the language `#1'. Using the pattern for}%
\typeout{** the default language instead.}%
\else
\language=\csname l@#1\endcsname
\fi
#2}}
\providecommand{\BIBdecl}{\relax}
\BIBdecl

\bibitem{KV:96}
H.~Karim and M.~Viberg, ``Two decades of array signal processing research: The
  parametric approach,'' \emph{IEEE Signal Processing Magazine}, vol.~13,
  no.~4, Jul. 1996.

\bibitem{candes2008}
E.~J. Cand{\`e}s and M.~B. Wakin, ``An introduction to compressive sampling,''
  \emph{IEEE Signal Processing Magazine}, vol.~25, no.~2, pp. 21--30, 2008.

\bibitem{MCW:05}
D.~Malioutov, M.~Cetin, and A.~S. Willsky, ``A sparse signal reconstruction
  perspective for source localization with sensor arrays,'' \emph{IEEE
  Transactions on Signal Processing}, vol.~53, no.~8, pp. 3010--3022, Aug.
  2005.

\bibitem{HM:10}
M.~Hyder and K.~Mahata, ``Direction-of-arrival estimation using a mixed
  $\ell_{2,0}$ norm approximation,'' \emph{IEEE Transactions on Signal
  Processing}, vol.~58, no.~9, pp. 4646--4655, Sep. 2010.

\bibitem{ZLG:11}
H.~Zhu, G.~Leus, and G.~B. Giannakis, ``Sparsity-cognizant total least-squares
  for perturbed compressive sampling,'' \emph{IEEE Transactions on Signal
  Processing}, vol.~59, no.~5, May 2011.

\bibitem{SRHD:18}
S.~Semper, F.~Römer, T.~Hotz, and G.~{Del Galdo}, ``{Grid-Free}
  {Direction-of-Arrival} estimation with compressed sensing and arbitrary
  antenna arrays,'' in \emph{Proceedings of the IEEE International Conference
  on Acoustics, Speech, and Signal Processing (ICASSP 2018)}, Calgary, Canada,
  Apr. 2018.

\bibitem{heckel2017gen_line_spec_est}
R.~Heckel and M.~Soltanolkotabi, ``Generalized line spectral estimation via
  convex optimization,'' \emph{IEEE Transactions on Information Theory},
  vol.~PP, no.~99, pp. 1--1, 2017.

\bibitem{boyd2011admm}
S.~Boyd, N.~Parikh, E.~Chu, B.~Peleato, and J.~Eckstein, ``Distributed
  optimization and statistical learning via the alternating direction method of
  multipliers,'' \emph{Foundations and Trends® in Machine Learning}, vol.~3,
  no.~1, pp. 1--122, 2011.

\bibitem{li2016offgridlinespectrumdenoising}
Y.~Li and Y.~Chi, ``Off-the-grid line spectrum denoising and estimation with
  multiple measurement vectors,'' \emph{IEEE Transactions on Signal
  Processing}, vol.~64, no.~5, pp. 1257--1269, March 2016.

\bibitem{bhaskar2013ANDNlinespecest}
B.~N. Bhaskar, G.~Tang, and B.~Recht, ``Atomic norm denoising with applications
  to line spectral estimation,'' \emph{IEEE Transactions on Signal Processing},
  vol.~61, no.~23, pp. 5987--5999, 2013.

\bibitem{LRT:04}
M.~Landmann, A.~Richter, and R.~S. Thomä, ``{DoA} resolution limits in {MIMO}
  channel sounding,'' in \emph{IEEE Antennas and Propagation Society
  Symposium}, vol.~2, Jun. 2004, pp. 1708--1711.

\bibitem{angel2015comp_beamform}
\BIBentryALTinterwordspacing
A.~Xenaki and P.~Gerstoft, ``Grid-free compressive beamforming,'' \emph{The
  Journal of the Acoustical Society of America}, vol. 137, no.~4, pp.
  1923--1935, 2015. [Online]. Available:
  \url{https://doi.org/10.1121/1.4916269}
\BIBentrySTDinterwordspacing

\bibitem{TBSR:13}
G.~Tang, B.~N. Bhaskar, P.~Shah, and B.~Recht, ``Compressed sensing off the
  grid,'' \emph{IEEE Transactions on Information Theory}, vol.~59, no.~11, pp.
  7465--7490, Nov. 2013.

\bibitem{lichi2016admm_mmv}
Y.~Li and Y.~Chi, ``Off-the-grid line spectrum denoising and estimation with
  multiple measurement vectors,'' \emph{IEEE Transactions on Signal
  Processing}, vol.~64, no.~5, pp. 1257--1269, March 2016.

\bibitem{candes2013superresnoisydata}
\BIBentryALTinterwordspacing
E.~J. Cand{\`e}s and C.~Fernandez-Granda, ``Super-resolution from noisy data,''
  \emph{Journal of Fourier Analysis and Applications}, vol.~19, no.~6, pp.
  1229--1254, 2013. [Online]. Available:
  \url{http://dx.doi.org/10.1007/s00041-013-9292-3}
\BIBentrySTDinterwordspacing

\bibitem{yang2016MLtoeplitz}
Z.~Yang, L.~Xie, and P.~Stoica, ``Vandermonde decomposition of multilevel
  toeplitz matrices with application to multidimensional super-resolution,''
  \emph{IEEE Transactions on Information Theory}, vol.~62, no.~6, pp.
  3685--3701, June 2016.

\bibitem{IRLKRLGDT:17}
M.~Ibrahim, V.~Ramireddy, A.~Lavrenko, J.~K\"onig, F.~R\"omer, M.~Landmann,
  M.~Grossmann, G.~D. Galdo, and R.~S. Thomä, ``Design and analysis of
  compressive antenna arrays for direction of arrival estimation,''
  \emph{Elsevier Signal Processing}, vol. 138, pp. 35 -- 47, Sep. 2017.

\bibitem{roy1989esprit}
R.~Roy and T.~Kailath, ``Esprit-estimation of signal parameters via rotational
  invariance techniques,'' \emph{IEEE Transactions on Acoustics, Speech, and
  Signal Processing}, vol.~37, no.~7, pp. 984--995, July 1989.

\end{thebibliography}
